\documentclass[preprint,prd,aps,tighten,nofootinbib,amssymb]{revtex4}

\usepackage{graphicx}
\usepackage{epsfig,amssymb,amsmath}

\pdfoutput=1
\usepackage{subfigure}
\usepackage[colorlinks=true,linkcolor=blue,citecolor=blue]{hyperref}
\usepackage{float}

\makeatletter
\def\Hline{%
\noalign{\ifnum0=`}\fi\hrule \@height 2pt \futurelet
\reserved@a\@xhline}
\makeatother

\newcommand{\beq}{\begin{equation}}
\newcommand{\eeq}{\end{equation}}
\newcommand{\bea}{\begin{eqnarray}}
\newcommand{\eea}{\end{eqnarray}}
\newcommand{\bear}{\begin{array}}
\newcommand {\eear}{\end{array}}
\newcommand{\bef}{\begin{figure}}
\newcommand {\eef}{\end{figure}}
\newcommand{\bec}{\begin{center}}
\newcommand {\eec}{\end{center}}

\def\EQ#1{Eq.~(\ref{#1})}

\def\REF#1{(\ref{#1})}
\def\GEV#1{10^{#1}{\rm\,GeV}}

\def\lrfp#1#2#3{ \left(\frac{#1}{#2} \right)^{#3}}

\begin{document}
\draft
\tighten
\preprint{TU-991,~IPMU15-0016}
\title{\large \bf
Gravitational waves from Higgs domain walls
}
\author{
Naoya Kitajima\,$^{a}$\footnote{email:kitajima@tuhep.phys.tohoku.ac.jp},
Fuminobu  Takahashi\,$^{a,b}$\footnote{email: fumi@tuhep.phys.tohoku.ac.jp}
}
\affiliation{
$^a$ Department of Physics, Tohoku University, Sendai 980-8578, Japan\\
$^b$ Kavli IPMU, TODIAS, University of Tokyo, Kashiwa 277-8583, Japan
}

\vspace{2cm}

\begin{abstract}
The effective potential for the Standard Model Higgs field allows two quasi-degenerate vacua; one is
our vacuum at the electroweak scale, while the other is at a much higher scale. The latter minimum
may be at a scale much smaller than the Planck scale, if the potential is lifted by new physics. 
This gives rise to a possibility of domain wall formation after inflation. If the high-scale minimum is 
a local minimum, domain walls are unstable and disappear through violent annihilation processes,
producing  a significant amount of gravitational waves.
We estimate the amount of gravitational waves produced from unstable domain walls in the Higgs potential
and discuss detectability with  future experiments. 
\end{abstract}

\maketitle

\section{Introduction}
\label{sec:intro}

The Standard Model (SM) of particle physics has been extremely successful, and the Higgs boson with a mass of
$125$\,GeV
discovered at the LHC~\cite{Aad:2012tfa,Chatrchyan:2012ufa} completed the last missing piece of the SM. 
So far there is no experimental hint for physics beyond the SM, and  it may be that the Standard Model is valid up to very
high energy scales  beyond the reach of the current collider experiments.

In the SM framework, the measured values of the Higgs boson mass and top quark mass 
imply that the electroweak (EW) vacuum is likely metastable. This is because the Higgs self coupling becomes negative at 
some high energy scale as a result of the fact that its RGE (renormalization group equation) evolution is dominated by the top Yukawa 
coupling~\cite{Buttazzo:2013uya,Andreassen:2014gha}.
While such effective potential is acceptable as long as our vacuum is sufficiently long-lived,  it may
signal that new physics appears around that scale and lift the potential.\footnote{
The scale sensitively depends on the top quark mass, and it will be around or beyond the Planck scale 
for the top quark mass about $171$\,GeV, the lower side of the experimental range~\cite{Buttazzo:2013uya,Andreassen:2014gha}. }
For example, higher dimensional operators may lift the effective potential so as to make these two vacua degenerate in energy, or even make 
the EW vacuum stable. 

The existence of two quasi-degenerate minima in the Higgs potential has interesting cosmological implications. 
During inflation, either of the two vacua  is
randomly selected in each patch of the Universe, if
the Higgs field acquires quantum fluctuations large enough to overcome the potential barrier between the two minima.
Then, domain walls are formed after inflation.  Domain walls are a sheet-like topological defect \cite{vilenkin2000cosmic},
and they are stable if the two vacua are exactly degenerate. Stable domain walls, however, are a cosmological catastrophe
as they eventually dominate the Universe, generating unacceptably large inhomogeneities.\footnote{
If the high-scale minimum is around or beyond the Planck scale, an eternal topological Higgs inflation may take place
avoiding the cosmological disaster~\cite{Hamada:2014raa}. See Refs.~\cite{Linde:1994hy,Vilenkin:1994pv} for
the original works of topological inflation. The topological Higgs inflation may provide a dynamical explanation for
the multiple-point criticality principle \cite{Froggatt:1995rt}. See also Ref.~\cite{Hamada:2015ria} for the related topics.}
If there is an energy difference (bias) between the two vacua, 
domain walls are unstable and they eventually annihilate~\cite{Vilenkin:1981zs,Gelmini:1988sf,Coulson:1995nv,Larsson:1996sp}. The EW vacua is realized
in the whole Universe if it is energetically preferred. Interestingly,
a significant amount of gravitational waves is emitted in the violent 
annihilation processes~\cite{Gleiser:1998na,Hiramatsu:2010yz,Kawasaki:2011vv,Hiramatsu:2013qaa}.  As we shall see shortly, 
such gravitational waves are within the sensitivity reach of  future experiments such as advanced-LIGO \cite{Abramovici:1992ah}, 
KAGRA \cite{Somiya:2011np,Aso:2013eba}, ET \cite{Sathyaprakash:2012jk}, LISA \cite{AmaroSeoane:2012km} and DECIGO \cite{Kawamura:2006up},
if the domain walls are sufficiently long-lived. Thus, gravitational waves can be a probe of another vacuum far beyond the EW scale.\footnote{
The gravitational waves from domain wall annihilation can also be a probe of the SUSY breaking scale~\cite{Takahashi:2008mu} or
a thermal inflation scenario~\cite{Moroi:2011be}.
}

In this letter, we study the gravitational waves generated by collapsing domain walls in the Higgs potential with the quasi-degenerate vacua.
In Sec.~\ref{sec:higgs}, we introduce the Higgs potential having the false vacuum at high energy scales and discuss the possibility of the domain wall 
formation. In Sec.~\ref{sec:gw}, we calculate the gravitational wave abundance and discuss its detectability with future experiments.
 Sec.~\ref{sec:conc} is devoted to discussion  and conclusions.

\section{False vacuum in Higgs potential}
\label{sec:higgs}

Let us consider the following Higgs potential lifted by new physics at some high energy scale,
\beq
	V_H = \frac{1}{4} \lambda(\varphi) \varphi^4 + \frac{\varphi^6}{\Lambda^2}.
	\label{eq:higgs_pot}
\eeq
where $\varphi$ is the Standard Model Higgs scalar field, $\lambda(\varphi)$ is a scale-dependent self coupling constant and $\Lambda$ is a cutoff scale
for the dimension six operator.
We have neglected the quadratic term of order  the electroweak scale as we are interested in the behavior of the Higgs potential at high energy.
Here we simply substitute the Higgs field value for the renormalization scale, $\lambda(\mu) = \lambda(\varphi)$.

The  renormalization-group-improved effective potential including one-loop and two-loop corrections in Landau gauge
is given in \cite{Buttazzo:2013uya}, and the gauge-dependence of the
effective potential was examined in \cite{Andreassen:2014gha}. More recently, a treatment for  higher dimensional operators 
in the Higgs potential was studied in \cite{Eichhorn:2015kea}. For our order of magnitude estimate, however, the following crude approximation is 
sufficient. We leave further refinement of our analysis for future work, but we believe our main results will not be qualitatively changed.

The scale dependence of the Higgs self coupling $\lambda(\mu)$ is governed by the RGE, 
\beq
	(4\pi)^2 \frac{d\lambda}{dt} = \beta_\lambda,
\eeq
where $t = \ln(\mu/M_t)$ and 
\beq
	\beta_\lambda = 24 \lambda^2 + 12 \lambda y_t^2 -6y_t^4-3\lambda (g'^2+3g^2)+\frac{3}{8}[2g^4+(g'^2+g^2)^2]
	\label{beta_lambda}
\eeq
up to one loop order.  Here $y_t$ is the top Yukawa coupling, $M_t$ is the top quark mass, 
$g'$ and $g$ are respectively the ${\rm U(1)_Y}$ and the ${\rm SU(2)_L}$ gauge coupling constants.
 The top Yukawa coupling gives a dominant contribution to the RGE up to a very high energy scale where the U(1)$_Y$ gauge
 coupling becomes large. As a result, the Higgs self coupling turns to negative, and the effective potential becomes
 negative at an intermediate scale in the absence of  higher dimensional operators.

The effective potential can be lifted by higher dimensional operators such as
the last term in the right-hand-side in Eq.~(\ref{eq:higgs_pot}).\footnote{
Alternatively, one may lift the potential by introducing e.g. an additional singlet scalar field which 
gives a positive contribution to the beta function  (\ref{beta_lambda}) \cite{Haba:2013lga,Khan:2014kba,Kawana:2014zxa}.
Our results hold in this case too, without significant modifications. 
}
 In this case there are two potential minima; one is at the EW scale $\varphi = v_{\rm EW} $, and the other 
at a much higher scale $\varphi = \varphi_f$. Depending on the size of the higher dimensional operator, the high-scale minimum can be a local or
global minimum. In particular, our main interest lies in the case when the two minima are quasi-degenerate and the EW vacuum is slightly
energetically preferred:
\bea
&V_H(v_{\rm EW}) \approx V_H(\varphi_f) \approx 0, \\
&V_H(v_{\rm EW}) <V_H(\varphi_f).
\eea
If the Higgs field acquires a sufficiently large quantum fluctuations during inflation,
 both vacua may be populated in different patches of the Universe, leading to domain wall formation 
after inflation. In a later Universe,  the EW vacuum will be selected after domain wall annihilation. 

\begin{figure}[tp]
\centering
\includegraphics [width = 10cm, clip]{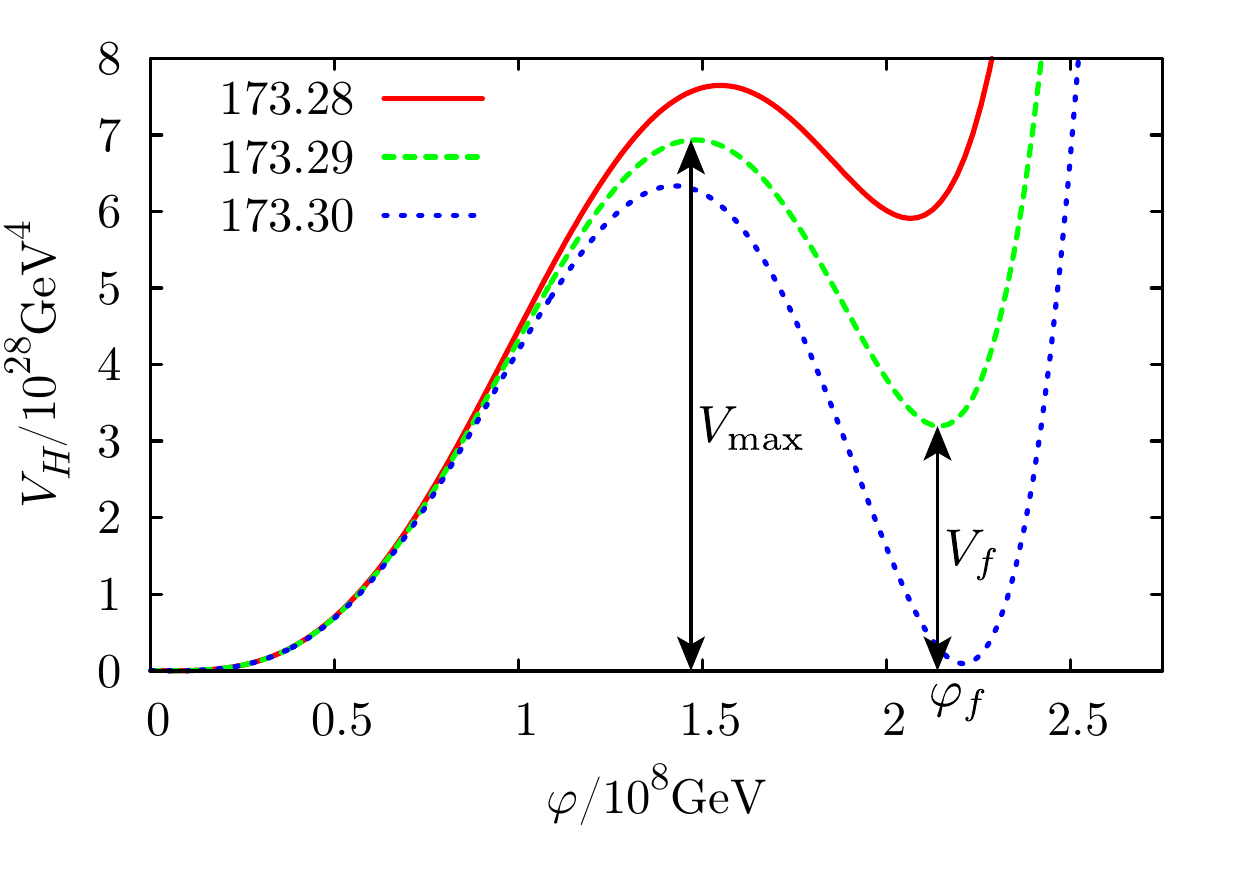}
\caption{
	The effective potential for the Higgs field, $V_H(\varphi)$.
	We have taken $M_t =$ 173.28 (solid red), 173.29 (dashed green) and 173.30 GeV (dotted blue), 
	$m_H = $ 125.6 GeV, and  $\Lambda = 10^{10}$ GeV. Note that the values of $\Lambda$ and $\varphi_f$
	are  for illustration purposes only. See the text for details. 
}
\label{fig:higgs_pot}
\end{figure}

In Fig.~\ref{fig:higgs_pot} we show the Higgs potential calculated in the present set-up  for illustration purposes.
We have taken the Higgs boson mass $m_H = 125.6$ GeV, $M_t = 173.28$ (solid (red)), 173.29 (dashed (green)), 
173.30 GeV(dotted (blue)) and $\Lambda = 10^{10}$ GeV. One can see that the position of the high-scale minimum, $\varphi = \varphi_f$, is
comparable to, but parametrically smaller than $\Lambda$ because of the loop suppression factor in the RGE.
The cut-off scale as well as $\varphi_f$ increase for smaller values of $M_t$.  
For notational simplicity, let $V_f$ and $V_{\rm max}$ denote the potential energy at the false vacuum and the local maximum,
respectively, as shown in the figure for the middle line. 

Note that the precise value of $\Lambda$ obtained in Ref.~\cite{Andreassen:2014gha}
is about two orders of magnitude larger and it is about $\GEV{12}$ for $M_t =173.3$\,GeV. Accordingly, $\varphi_f$
will be larger by a similar amount. We emphasize here that our analysis in the next section
does not depend on the detailed RGE evolution, because we express all the relevant
quantities in terms of $\varphi_f$, $V_{\rm max}$ and $V_f$. One should simply use the result of
e.g. Ref.~\cite{Andreassen:2014gha} when one relates the value of $\varphi_f$ to the
top quark mass and the Higgs boson mass.

\section{Gravitational waves from collapsing domain walls}
\label{sec:gw}
As we have seen in the previous section, the Higgs potential allows two quasi-degenerate minima, especially 
if the potential is lifted by higher dimensional operators. This gives rise to a possibility of domain wall formation after inflation.
A domain wall is characterized by its tension, $\sigma$, which is roughly estimated to be
\beq
	\sigma \sim \bigg( \frac{\varphi_f^2}{w^2} + V_{\rm max} \bigg) w \sim V_{\rm max}^{1/2} \varphi_f,
	\label{sigma}
\eeq
where  $V_{\rm max}$ is the height of the potential barrier between two minima, 
$w \sim \varphi_f/V_{\rm max}^{1/2}$ is the width of the domain wall, and
we fix it to minimize the tension in the second equality.
In order to avoid the cosmological domain wall problem, the energy bias between the two vacua is necessary to make domain walls unstable.
In the presence of the bias, domain walls  start to collapse when the energy density of the domain walls become comparable to the bias energy density. 
As is confirmed by numerical simulations, the evolution of the domain wall network exhibits a scaling behavior \cite{Press:1989yh,Hindmarsh:1996xv,Garagounis:2002kt,Leite:2011sc} and the energy density of the domain wall is roughly given by 
\beq
\rho_{\rm dw} \sim \sigma H.
\label{rhodw}
\eeq
Then, the Hubble parameter at the domain wall decay is 
\beq
	H_{\rm dec} \sim \frac{V_f}{\sigma} \sim \frac{V_f}{V_{\rm max}^{1/2} \varphi_f}.
\eeq
In order not to generate unacceptably large  inhomogeneities,  domain walls must decay before they dominate the Universe,
which places a lower bound on the energy bias;
\bea
H_{\rm dec} &>& H_{\rm dom} ~~\Longleftrightarrow~~ \frac{V_f}{V_{\rm max}} > \lrfp{\varphi_f}{M_P}{2},
\label{DWdom}
\eea
where $M_P \simeq 2.4 \times \GEV{18}$ is the reduced Planck mass, and 
$H_{\rm dom} \sim \sigma/M_P^2$ is the Hubble parameter when domain walls would start to dominate the energy density of the Universe if there were not
for the bias.

The domain wall collapse is a violent processes, and some part of the energy stored in the domain walls is 
converted to gravitational waves. The spectrum of the gravitational waves is expected to be
peaked at a frequency  corresponding to the Hubble scale at the decay, as it is the typical curvature scale of
the domain wall system.
This was confirmed by
detailed numerical calculations~\cite{Hiramatsu:2013qaa},  and
the density parameter of the gravitational waves at the peak frequency at the time of domain wall collapse is given by
\beq
	\Omega_{\rm GW}(t_{\rm dec})|_{\rm peak} = \frac{8\pi \tilde{\epsilon}_{\rm gw} G^2 \mathcal{A}^2 \sigma^2}{3 H_{\rm dec}^2} 
	= \frac{\tilde{\epsilon}_{\rm gw} \mathcal{A}^2}{24 \pi} \bigg( \frac{V_{\rm max}}{V_f} \bigg)^2 \bigg( \frac{\varphi_f}{M_P} \bigg)^4,
	\label{OGW}
\eeq
where $G$ is the Newton's gravitational constant and $\tilde{\epsilon}_{\rm gw}$ and $\mathcal{A}$ are numerical factors
 characterizing respectively the efficiency of the gravitational wave emission and the area of the domain walls. They are determined by the numerical calculations to be $\tilde{\epsilon}_{\rm gw} \simeq 0.7$ and $\mathcal{A} \simeq 0.8$ \cite{Hiramatsu:2013qaa}.
Then, the present time density parameter of the gravitational waves at the peak frequency is obtained as
\beq
	\Omega_{\rm GW} h^2|_{\rm peak} \simeq 1.3 \times 10^{-5} \gamma \bigg( \frac{106.75}{g_*} \bigg)^{1/3} \Omega_{\rm GW}(t_{\rm dec})|_{\rm peak} 
\eeq
where $g_*$ is the relativistic degrees of freedom at the domain wall decay and $\gamma$ is the dilution factor after the domain wall decay, and it is given by
\beq
\gamma \simeq \left\{
\bear{lc}
1 & ~~{\rm for}~~H_{\rm dec} < H_R\\
 (H_R/H_{\rm dec})^{2/3}& ~~{\rm for}~~H_{\rm dec} > H_R\\
\eear
\right..
\eeq
Here $H_R$ is the Hubble parameter at the reheating.
The peak frequency corresponds to the Hubble parameter at domain wall decay, which is red-shifted by the cosmic expansion until today,
\beq
	f_{\rm peak} = \frac{a(t_{\rm dec})}{a(t_0)} H_{\rm dec} \simeq 160 \,{\rm Hz} ~ \gamma^{-1/2} \bigg(\frac{g_*}{106.75} \bigg)^{1/6} \bigg(\frac{T_X}{10^9~{\rm GeV}} \bigg)
\eeq
where 
\beq
T_X \simeq \left\{
\bear{lc}
T_{\rm dec} & ~~{\rm for}~~H_{\rm dec} < H_R\\
T_R & ~~{\rm for}~~H_{\rm dec} > H_R\\
\eear
\right.,
\eeq
and $T_{\rm dec}$ and $T_R$ are the cosmic temperature at $H = H_{\rm dec}$ and $H_R$, respectively.

Now let us turn to the domain walls in the SM Higgs potential lifted by new physics. We focus on the case in which the high-scale minimum is a false vacuum and it is
quasi-degenerate with the EW vacuum. 
The position of the false vacuum is at intermediate energy scales, $10^8~{\rm GeV} \lesssim \varphi_f \lesssim 10^{12}~{\rm GeV}$, depending on the
values of the top quark mass, the strong gauge coupling, and the Higgs boson mass~\cite{Buttazzo:2013uya,Andreassen:2014gha}.
 The height of the potential barrier is determined by solving the RGE,
which is roughly $V_{\rm max} \sim 10^{-4} \varphi_f^4$.
The bias energy density is treated as a  free parameter which is adjusted by tuning $\Lambda$.

In the case of the Higgs domain walls, there are generically finite temperature corrections to the Higgs potential, which give an extra
contribution to the (time-dependent) energy bias. In the following we 
 derive a condition for thermal effects to have a negligible impact on the domain wall dynamics. 
The condition can be relaxed in certain situations, and we shall give concrete examples later.

Throughout this letter, we focus on the case in which  the position of the false vacuum is always much larger than the cosmic temperature,
$\varphi_f \gg T$.
Then the thermal mass correction to the effective potential is negligibly small at $\varphi = \varphi_f$. Even for $\varphi \gg T$, however, 
there is a logarithmic correction arising  from the free energy of thermal plasma, the so called {\it thermal log} potential~\cite{Anisimov:2000wx},
 which is roughly given by
\beq
	V_T(\varphi) = a T^4 \ln \bigg(\frac{\varphi}{T} \bigg),
\eeq
where $a$ is a numerical constant of $O(0.1)$. In addition, there are background thermal plasma in the EW vacuum, 
while many of the SM particles are non-relativistic in the false vacuum because of $\varphi_f \gg T$.  These thermal effects
are considered to generate an extra energy bias  of order $T^4$.

After reheating,  the thermal energy bias $\sim T^4$
decreases faster than the energy density of domain walls in the scaling regime. Before  reheating, both evolve in the same way
 since  the dilute plasma energy density evolves as $T^4 \sim T_R^2 H M_P  \propto \rho_{\rm dw}$ (see \EQ{rhodw})
 in the case of usual perturbative decay of the inflaton. Thermal effects do not induce the domain wall annihilation
if $\rho_{\rm dw} \sim \sigma H \gtrsim  T^4$.
For a given reheating temperature, this gives a lower bound on the field value at the false vacuum:
\beq
	T_R \lesssim 3 \times 10^8~{\rm GeV} \bigg( \frac{\varphi_f}{10^{12}~{\rm GeV}} \bigg)^{3/2} \lesssim V_f^\frac{1}{4},
	\label{lowTR}
\eeq
where the second inequality represents the condition for avoiding the domain wall domination (\ref{DWdom}).
The condition (\ref{lowTR}) implies that the domain walls must annihilate before the reheating.

For the reheating temperature satisfying the condition (\ref{lowTR}), the thermal plasma energy is always much smaller
than $V_{\rm max}$, and so, the use of the tension given in \EQ{sigma} is justified.\footnote{
The evolution of the domain wall network might deviate from the scaling regime in the presence of 
background plasma. The existence of plasma could also change
 the effective area of domain walls ${\cal A}$, which results in a shift of the peak frequency for the fixed tension and energy bias.
We however expect that such thermal effect on the domain wall dynamics is small as long as thermal energy is much smaller 
than the energy stored in the domain walls, as we assume in the text. 
} 
If the condition (\ref{lowTR}) is violated, domain walls will disappear soon after inflation
and the resultant gravitational wave signal will be too weak to be detected by future experiments.
Also the peak frequency tends to be extremely high.

We would like to emphasize that the bound (\ref{lowTR}) should be taken with care, because the thermal 
history after inflation is unknown. Indeed, it is possible to consider some complicated (and contrived) thermal 
history where thermal effects on the domain wall dynamics are negligible even if  (\ref{lowTR})  is not satisfied. 
For instance, the inflaton may dominantly decay into hidden sector particles, and the SM sector is reheated 
when the coupling between the SM and hidden sectors freezes-in well after the domain wall annihilation.\footnote{
Our estimate  (\ref{OGW}) remains unchanged in this case.} We shall return to this issue in Sec.~\ref{sec:conc}.

\begin{figure}[tp]
\centering
\subfigure[~$T_R = 3 \times 10^{8}$ GeV]{
\includegraphics [width = 10cm, clip]{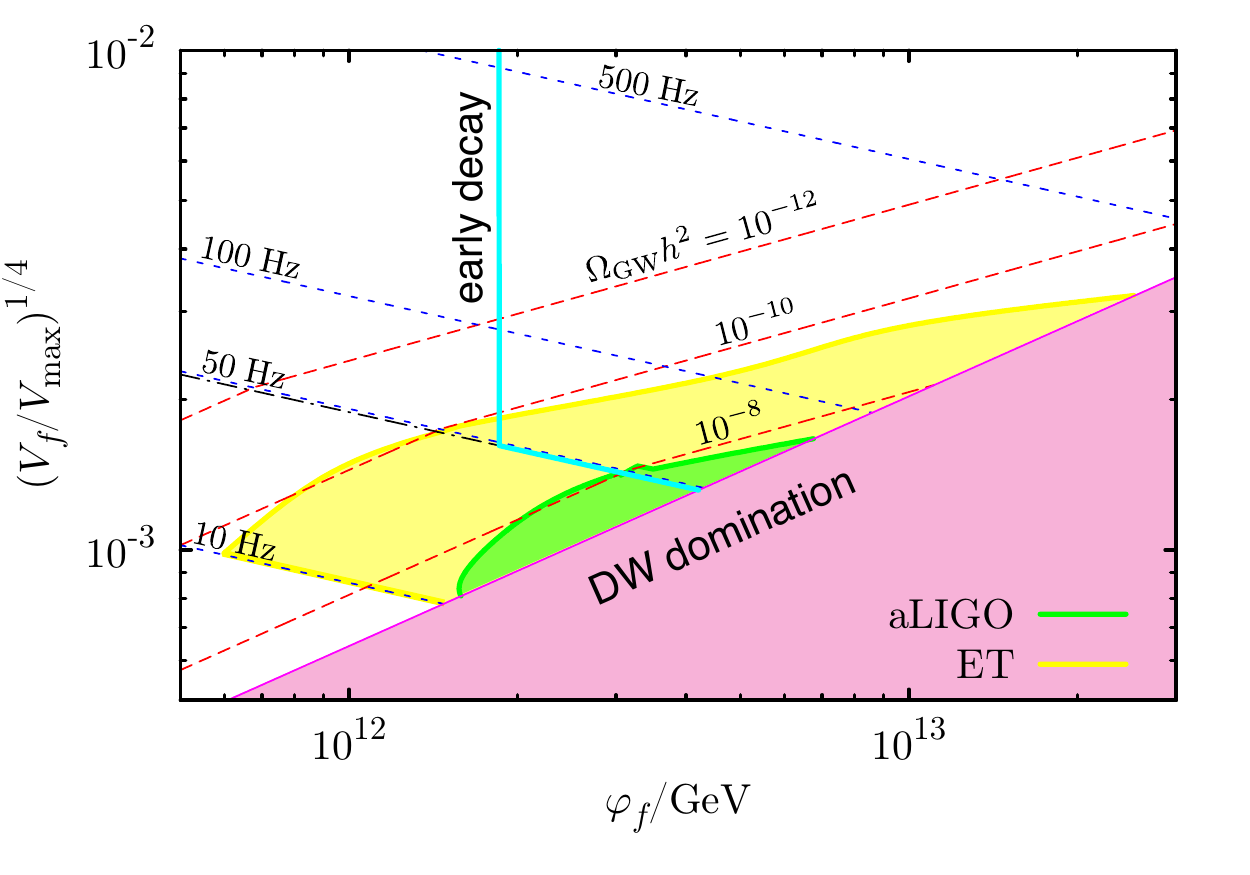}
\label{subfig:contour1}
}
\subfigure[~$T_R = 10^{4}$ GeV]{
\includegraphics [width = 10cm, clip]{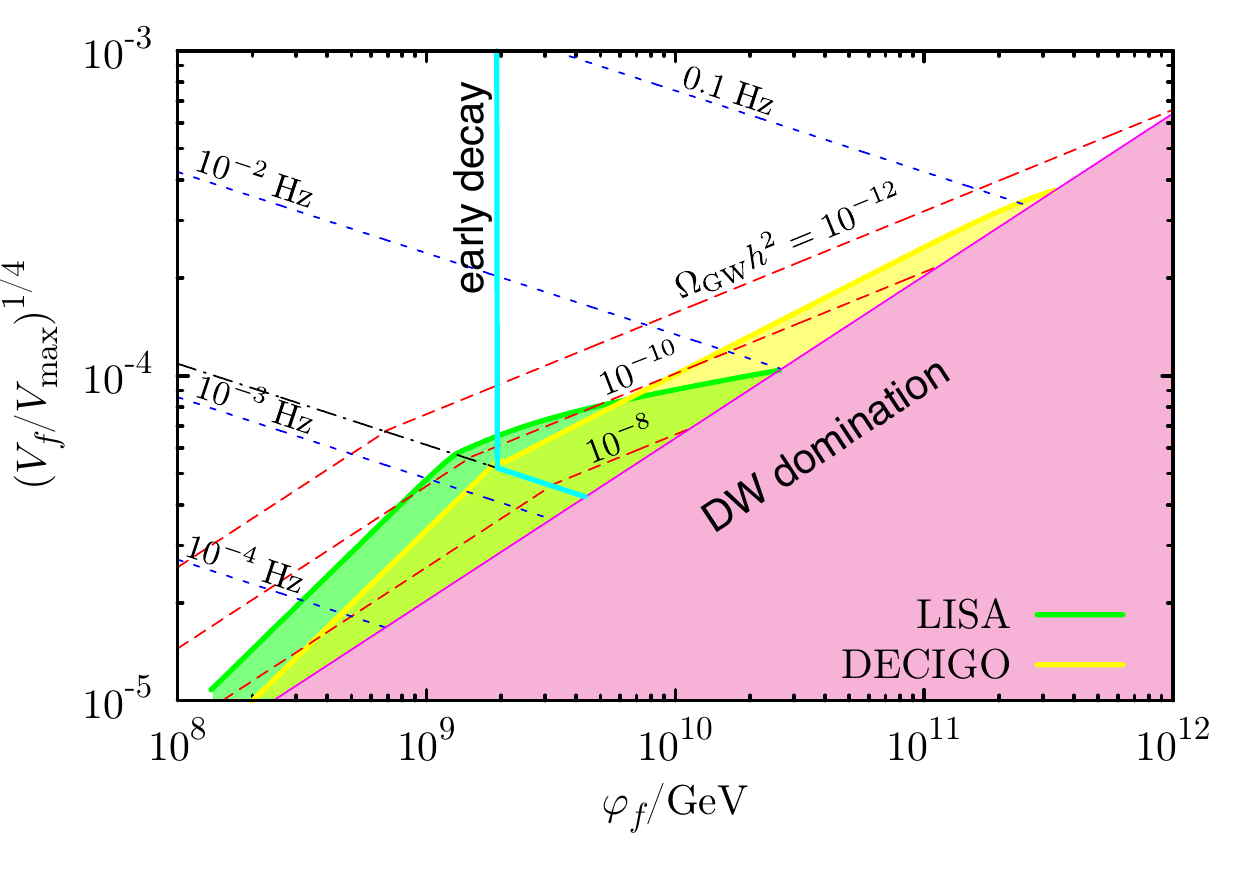}
\label{subfig:contour2}
}
\caption{
The contours of  $\Omega_{\rm GW}h^2|_{\rm peak}$ (dashed red) and $f_{\rm peak}$ (dotted blue) in 
the plane of $\varphi_f$--$(V_f/V_{\rm max})^{1/4}$.
We have set $T_R = 3 \times 10^8$ GeV and $T_R=10^{4}$ GeV in the upper and lower panel, respectively. 
The dash-dotted black line represents the border above (below) which the reheating takes place after (before) the wall decay \REF{lowTR}.
The lower right shaded (magenta) region is excluded by the domain wall domination  \REF{DWdom} 
and the solid cyan line corresponds to the lower bound on $\varphi_f$ due to the early domain wall decay by the thermal effect.
The thick green and yellow lines represent the sensitivity curves for advanced-LIGO (LISA) and ET (DECIGO) respectively in the upper (lower) panel
and the shaded region below the curves will be  probed by each  experiment.
}
\label{fig:contour}
\end{figure}

\begin{figure}[th]
\centering
\includegraphics [width = 10cm, clip]{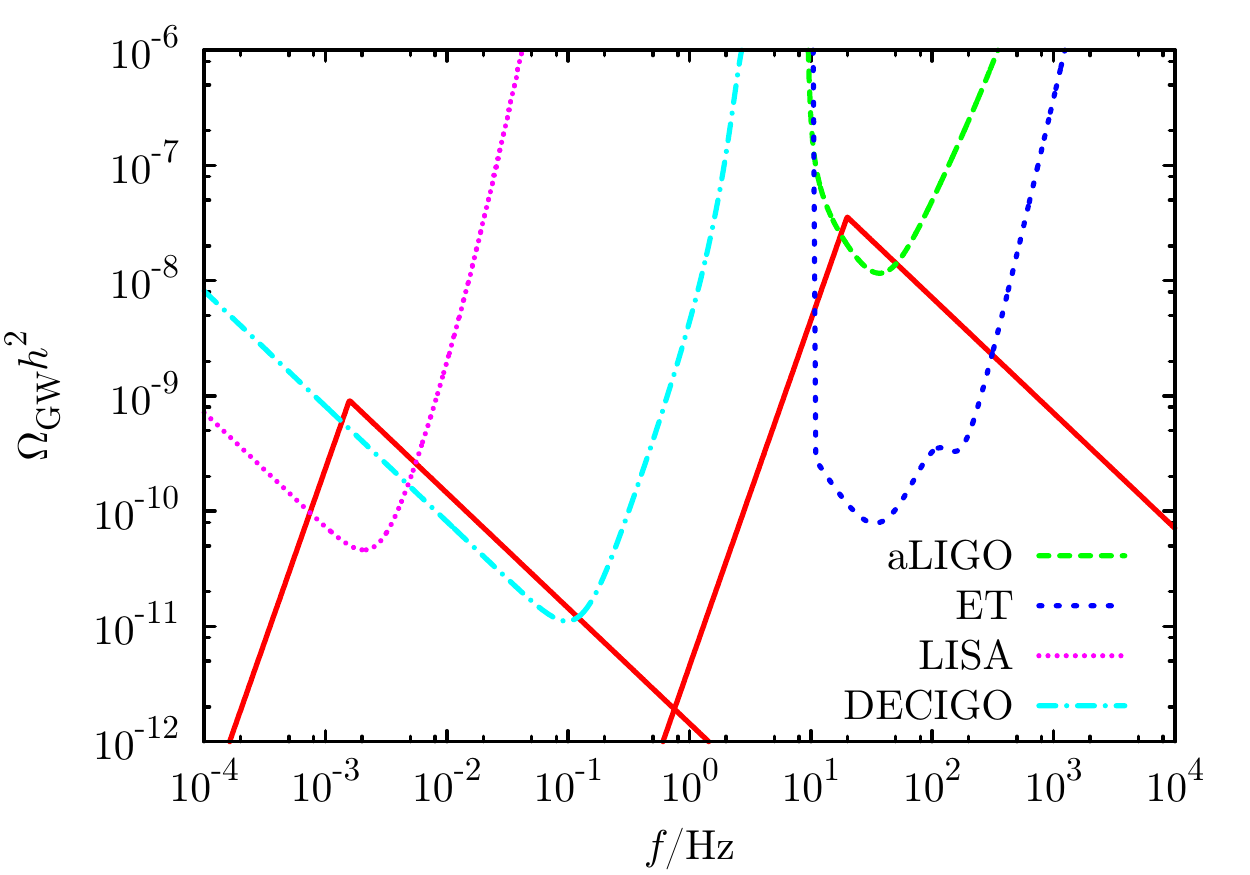}
\caption{
	The typical spectrum of the gravitational waves is shown by the solid (red) lines. 
	We have taken $\varphi_f = 2\times 10^{9}~{\rm GeV}$ and $(V_f/V_{\rm max})^{1/4} = 5 \times 10^{-5}$ for the left line and $\varphi_f = 2\times 10^{12}~{\rm GeV}$ and $(V_f/V_{\rm max})^{1/4} = 10^{-3}$ for the right line.
}
\label{fig:spectrum}
\end{figure}

In Fig.~\ref{fig:contour}, we show the contours of the gravitational wave density parameter at the peak frequency, 
$\Omega_{\rm GW}h^2|_{\rm peak}$ (dashed (red) lines), as well as  the peak frequency, 
$f_{\rm peak}$ (dotted (blue) lines),  in the plane of $(\varphi_f, (V_f/V_{\rm max})^{1/4})$.
We have set $T_R = 3 \times 10^8$ GeV in Fig.~\ref{subfig:contour1} and $T_R=10^{4}$ GeV in Fig.~\ref{subfig:contour2}.
The shaded (magenta) lower-right triangle region is excluded by the domain wall domination (cf. \EQ{DWdom}) and the solid cyan 
line is the lower bound on $\varphi_f$  to avoid  the early domain wall decay due to the thermal effects (cf. \EQ{lowTR}).
As mentioned earlier,  thermal effects are negligible in the region right to the solid cyan line, but it does not necessarily
preclude the region left to the solid cyan line because of large uncertainties of thermal history.
The thick green and yellow lines 
represent the sensitivity curves of advanced-LIGO and ET  respectively in Fig.~\ref{subfig:contour1},
and LISA and DECIGO in Fig.~(\ref{subfig:contour2}), and the shaded region below the curves will be probed by each 
experiments. The sensitivity curve of KAGRA
is expected to be similar to that of advanced-LIGO.
Note that it is difficult to generate gravitational waves within the reach of pulsar timing observations such as 
IPTA \cite{IPTA:2013lea} and SKA \cite{Kramer:2004rwa} which are sensitive to much lower frequencies ($\sim 10^{-9}$ Hz).
It would require the tension of the domain walls to be close to $(\GEV{5})^3$, which cannot be realized within the SM framework.

We show in Fig.~\ref{fig:spectrum} the gravitational wave spectrum for certain parameters ($\varphi_f = 2\times 10^{9}~{\rm GeV}$ and $(V_f/V_{\rm max})^{1/4} = 5 \times 10^{-5}$ for the left line and $\varphi_f = 2\times 10^{12}~{\rm GeV}$ and $(V_f/V_{\rm max})^{1/4} = 10^{-3}$ for the right line).
According to the precise numerical calculations \cite{Hiramatsu:2013qaa}, the gravitational wave  
spectrum scales as $(f/f_{\rm peak})^3$ for $f<f_{\rm peak}$ and $(f/f_{\rm peak})^{-1}$ for $f>f_{\rm peak}$.

\section{Discussion and Conclusions}
\label{sec:conc}
In the previous section we have mentioned that the condition (\ref{lowTR}) can 
be relaxed in certain situations. Here we give concrete examples 
that relax the condition (\ref{lowTR}) so as to make the wider parameter available. 
As we shall see below, this requires a more involved thermal history of the Universe.

Suppose that the inflaton dominantly decays into a hidden sector, which is decoupled from the SM sector. This significantly suppresses 
thermal corrections to the Higgs potential, and the lower bound on $\varphi_f$ (\ref{lowTR}) is no longer applied.\footnote{
Some amount of the SM plasma is necessarily generated by the Higgs domain walls, because they
continuously annihilate  in the scaling regime. We have numerically confirmed that the energy density
of the SM plasma induced by the Higgs domain walls in the scaling regime is always smaller than 
that of domain walls, and therefore, it does not induce the domain wall annihilation. } 
The SM particles can be thermally populated through various processes much later than the domain wall annihilation.
For instance, the hidden sector may contain a U(1) gauge group, which has a small kinetic mixing with 
the SM U(1)$_Y$.
For a sufficiently small kinetic mixing, the SM particles are thermalized through the kinetic mixing well after the domain wall annihilation
but before nucleosynthesis, unless the hidden photon is extremely light. If the hidden photon is massless, 
we need to introduce hidden matter
fields (e.g. a Dirac fermion) charged under the hidden U(1) gauge symmetry. The hidden matter fields behave
as mini-charged particles, connecting the hidden sector and the SM sector. In this case, the massless hidden photon contributes
to the effective neutrino species, $N_{\rm eff}$, whose precise value depends on the mass of the hidden matter fields~\cite{Vogel:2013raa}.
Similarly, if one introduces a gauge singlet scalar field, the SM sector can be produced from the hidden sector through the Higgs portal coupling.
In this case, the thermalization of the SM sector can be delayed for a sufficiently small Higgs portal coupling.
It is worth noting that there is no entropy production when the SM sector is thermalized, and our analysis in the previous
section can be applied without any modification even in the parameter space left to the solid cyan line in Fig.~\ref{fig:contour}.

Alternatively, the SM sector can be reheated, if one 
introduces another field (e.g. a modulus field) which dominates the Universe for a short period and decays into the SM particles 
after the inflaton decay. In this case one has to take account of its thermal effects, the shift of the peak frequency, as well as
an extra dilution due to the modified thermal history.

In the SM framework, the Higgs self coupling turns to
negative, and the effective potential becomes negative at a high energy scale, based on the perturbative RGE analysis.
The scale sensitively depends on the values of the top quark mass, but it is at an intermediate scale for the
top quark mass, $M_t \simeq 173$\,GeV. In such a case, the EW vacuum is metastable, which is acceptable as long
as it is sufficiently long-lived. On the other hand, the negative effective potential may signal that new physics appears
around that scale and lift the potential, creating a local minimum at an intermediate scale. In a limiting case, the two
vacua, one at the EW scale and the other at an intermediate scale, are quasi-degenerate in energy. This requires
 a certain amount of fine-tuning of the parameters. Once the fine-tuning of the parameters is realized,
it gives rise to a possibility of domain wall formation.
 
For domain walls to be formed after inflation, both vacua must be populated with more or less equal probability 
during inflation. Domain walls are not formed if one of the two vacua is preferred over the other. This is the case 
if  the Higgs potential is significantly modified through its non-minimal coupling to gravity, or if the quantum 
fluctuations of the Higgs field is too small to overcome the barrier separating the two vacua. We have assumed 
that the Higgs field is minimally coupled to gravity as well as the inflaton sector so that the potential is not significantly 
modified, and that the quantum fluctuations of the Higgs field is sufficiently large to overcome the barrier between
the two vacua. 

The energy of domain walls is partially converted to gravitational waves through the violent annihilation processes.
The gravitational waves are expected to be peaked at a frequency corresponding to the Hubble horizon size at the 
domain-wall annihilation.  The gravitational-wave spectrum looks like that from a phase transition. In the latter
case, the gravitational wave spectrum has a peak corresponding to the typical size of bubbles, a few orders of magnitude
smaller than the Hubble horizon at the phase transition. 

In order to generate gravitational waves within the reach of future experiments, the bias energy density must be so small
that domain walls annihilate when they are about to dominate the Universe. 
While there is no compelling reason for
the quasi-degeneracy of the two vacua, it may be due to the multiple-point principle~\cite{Froggatt:1995rt} or
an important cosmological role  (e.g. dark matter, baryogenesis) of the decay products of domain walls.
The violent domain wall annihilation processes produce not only gravitational waves but also
a large amount of the SM Higgs bosons which soon decay into quarks, leptons and gauge bosons.
If the SM Higgs boson is coupled to some heavy degrees of freedom such as right-handed neutrinos,
a B-L Higgs field, and a singlet scalar,  those heavy particles can be produced through non-perturbative processes, 
and they may contribute to dark matter, or baryogenesis, etc.~\cite{DKT}.

 In this letter we have investigated the domain wall formation 
in the Standard Model Higgs potential lifted by new physics and gravitational wave production due to the domain wall decay. 
We have shown that the gravitational waves can be within the reach of the future experiments  such as advanced LIGO, KAGRA,
ET, LISA and DECIGO, if the domain walls decay when their energy density is sizable.  
In this way, the direct detection experiments of the gravitational wave can be a powerful probe of another vacuum far beyond the EW scale.

\section*{Acknowledgment}
This work was supported by  JSPS Grant-in-Aid for
Young Scientists (B) (No.24740135 [FT]), 
Scientific Research (A) (No.26247042 [FT]), Scientific Research (B) (No.26287039 [FT]), 
 the Grant-in-Aid for Scientific Research on Innovative Areas (No.23104008 [NK, FT]),  and
Inoue Foundation for Science [FT].  This work was also
supported by World Premier International Center Initiative (WPI Program), MEXT, Japan [FT].


\begin{thebibliography}{99}

\bibitem{Aad:2012tfa} 
  G.~Aad {\it et al.}  [ATLAS Collaboration],
  Phys.\ Lett.\ B {\bf 716}, 1 (2012)
  [arXiv:1207.7214 [hep-ex]].

\bibitem{Chatrchyan:2012ufa} 
  S.~Chatrchyan {\it et al.}  [CMS Collaboration],
  Phys.\ Lett.\ B {\bf 716}, 30 (2012)
  [arXiv:1207.7235 [hep-ex]].
  
\bibitem{Buttazzo:2013uya} 
  D.~Buttazzo, G.~Degrassi, P.~P.~Giardino, G.~F.~Giudice, F.~Sala, A.~Salvio and A.~Strumia,
  JHEP {\bf 1312}, 089 (2013)
  [arXiv:1307.3536 [hep-ph]].

\bibitem{Andreassen:2014gha} 
  A.~Andreassen, W.~Frost and M.~D.~Schwartz,
  Phys.\ Rev.\ Lett.\  {\bf 113}, no. 24, 241801 (2014)
  [arXiv:1408.0292 [hep-ph]].
  
\bibitem{vilenkin2000cosmic}
  A.~Vilenkin and E. Shellard,
  {\it Cosmic Strings and Other Topological Defects.}
  Cambridge Monographs on Mathematical Physics.
  Cambridge University Press, 2000.
  
\bibitem{Hamada:2014raa} 
  Y.~Hamada, K.~y.~Oda and F.~Takahashi,
  Phys.\ Rev.\ D {\bf 90}, no. 9, 097301 (2014)
  [arXiv:1408.5556 [hep-ph]].
  
\bibitem{Linde:1994hy} 
  A.~D.~Linde,
  Phys.\ Lett.\ B {\bf 327}, 208 (1994)
  [astro-ph/9402031].

\bibitem{Vilenkin:1994pv} 
  A.~Vilenkin,
  Phys.\ Rev.\ Lett.\  {\bf 72}, 3137 (1994)
  [hep-th/9402085].
  
\bibitem{Froggatt:1995rt} 
  C.~D.~Froggatt and H.~B.~Nielsen,
  Phys.\ Lett.\ B {\bf 368}, 96 (1996)
  [hep-ph/9511371].

\bibitem{Hamada:2015ria} 
  Y.~Hamada, H.~Kawai and K.~y.~Oda,
  arXiv:1501.04455 [hep-ph].
  
\bibitem{Vilenkin:1981zs} 
  A.~Vilenkin,
  Phys.\ Rev.\ D {\bf 23}, 852 (1981).
  
\bibitem{Gelmini:1988sf} 
  G.~B.~Gelmini, M.~Gleiser and E.~W.~Kolb,
  Phys.\ Rev.\ D {\bf 39}, 1558 (1989).
  
\bibitem{Coulson:1995nv} 
  D.~Coulson, Z.~Lalak and B.~A.~Ovrut,
  Phys.\ Rev.\ D {\bf 53}, 4237 (1996).
  
\bibitem{Larsson:1996sp} 
  S.~E.~Larsson, S.~Sarkar and P.~L.~White,
  Phys.\ Rev.\ D {\bf 55}, 5129 (1997)
  [hep-ph/9608319].


\bibitem{Gleiser:1998na} 
  M.~Gleiser and R.~Roberts,
  Phys.\ Rev.\ Lett.\  {\bf 81}, 5497 (1998)
  [astro-ph/9807260].
  
\bibitem{Hiramatsu:2010yz} 
  T.~Hiramatsu, M.~Kawasaki and K.~Saikawa,
  JCAP {\bf 1005}, 032 (2010)
  [arXiv:1002.1555 [astro-ph.CO]].
  
\bibitem{Kawasaki:2011vv} 
  M.~Kawasaki and K.~Saikawa,
  JCAP {\bf 1109}, 008 (2011)
  [arXiv:1102.5628 [astro-ph.CO]].
  
\bibitem{Hiramatsu:2013qaa} 
  T.~Hiramatsu, M.~Kawasaki and K.~Saikawa,
  JCAP {\bf 1402}, 031 (2014)
  [arXiv:1309.5001 [astro-ph.CO]].


\bibitem{Abramovici:1992ah} 
  A.~Abramovici, W.~E.~Althouse, R.~W.~P.~Drever, Y.~Gursel, S.~Kawamura, F.~J.~Raab, D.~Shoemaker and L.~Sievers {\it et al.},
  Science {\bf 256}, 325 (1992).
  

\bibitem{Somiya:2011np} 
  K.~Somiya [KAGRA Collaboration],
  Class.\ Quant.\ Grav.\  {\bf 29}, 124007 (2012)
  [arXiv:1111.7185 [gr-qc]].
  
\bibitem{Aso:2013eba} 
  Y.~Aso {\it et al.}  [KAGRA Collaboration],
  Phys.\ Rev.\ D {\bf 88}, no. 4, 043007 (2013)
  [arXiv:1306.6747 [gr-qc]].

\bibitem{Sathyaprakash:2012jk} 
  B.~Sathyaprakash, M.~Abernathy, F.~Acernese, P.~Ajith, B.~Allen, P.~Amaro-Seoane, N.~Andersson and S.~Aoudia {\it et al.},
  Class.\ Quant.\ Grav.\  {\bf 29}, 124013 (2012)
  [Erratum-ibid.\  {\bf 30}, 079501 (2013)]
  [arXiv:1206.0331 [gr-qc]].

\bibitem{AmaroSeoane:2012km}
  P.~Amaro-Seoane, S.~Aoudia, S.~Babak, P.~Binetruy, E.~Berti, A.~Bohe, C.~Caprini and M.~Colpi {\it et al.},
  GW Notes {\bf 6} (2013) 4
  [arXiv:1201.3621 [astro-ph.CO]].

\bibitem{Kawamura:2006up} 
  S.~Kawamura, T.~Nakamura, M.~Ando, N.~Seto, K.~Tsubono, K.~Numata, R.~Takahashi and S.~Nagano {\it et al.},
  Class.\ Quant.\ Grav.\  {\bf 23}, S125 (2006).

\bibitem{Takahashi:2008mu} 
  F.~Takahashi, T.~T.~Yanagida and K.~Yonekura,
  Phys.\ Lett.\ B {\bf 664}, 194 (2008)
  [arXiv:0802.4335 [hep-ph]].

\bibitem{Moroi:2011be} 
  T.~Moroi and K.~Nakayama,
  Phys.\ Lett.\ B {\bf 703}, 160 (2011)
  [arXiv:1105.6216 [hep-ph]].

\bibitem{Eichhorn:2015kea} 
  A.~Eichhorn, H.~Gies, J.~Jaeckel, T.~Plehn, M.~M.~Scherer and R.~Sondenheimer,
  arXiv:1501.02812 [hep-ph].
  
  
  
\bibitem{ATLAS:2014wva} 
  [ATLAS and CDF and CMS and D0 Collaborations],
  arXiv:1403.4427 [hep-ex].
  
  
\bibitem{Haba:2013lga} 
  N.~Haba, K.~Kaneta and R.~Takahashi,
  JHEP {\bf 1404}, 029 (2014)
  [arXiv:1312.2089 [hep-ph]].
  
\bibitem{Khan:2014kba} 
  N.~Khan and S.~Rakshit,
  Phys.\ Rev.\ D {\bf 90}, no. 11, 113008 (2014)
  [arXiv:1407.6015 [hep-ph]].
  
\bibitem{Kawana:2014zxa} 
  K.~Kawana,
  arXiv:1411.2097 [hep-ph].


  
  
\bibitem{Press:1989yh} 
  W.~H.~Press, B.~S.~Ryden and D.~N.~Spergel,
  Astrophys.\ J.\  {\bf 347}, 590 (1989).
  
\bibitem{Hindmarsh:1996xv} 
  M.~Hindmarsh,
  Phys.\ Rev.\ Lett.\  {\bf 77}, 4495 (1996)
  [hep-ph/9605332].
  
\bibitem{Garagounis:2002kt} 
  T.~Garagounis and M.~Hindmarsh,
  Phys.\ Rev.\ D {\bf 68}, 103506 (2003)
  [hep-ph/0212359].
  
\bibitem{Leite:2011sc} 
  A.~M.~M.~Leite and C.~J.~A.~P.~Martins,
  Phys.\ Rev.\ D {\bf 84}, 103523 (2011)
  [arXiv:1110.3486 [hep-ph]].


\bibitem{Anisimov:2000wx} 
  A.~Anisimov and M.~Dine,
  Nucl.\ Phys.\ B {\bf 619}, 729 (2001)
  [hep-ph/0008058].
  
\bibitem{IPTA:2013lea} 
  R.~N.~Manchester,
  Class.\ Quant.\ Grav.\  {\bf 30}, 224010 (2013).
  
\bibitem{Kramer:2004rwa} 
  M.~Kramer, 2,
  astro-ph/0409020.
  
\bibitem{Vogel:2013raa} 
  H.~Vogel and J.~Redondo,
  JCAP {\bf 1402}, 029 (2014)
  [arXiv:1311.2600 [hep-ph]].
  
\bibitem{DKT}
R.~Daido, N.~Kitajima, F.~Takahashi, in preparation. 






\end{thebibliography}
\end{document}